\documentclass{PHYEAUTH}

\usepackage{graphicx}          
\usepackage{amsmath}           
\usepackage{amssymb}           

\begin{document}

\begin{frontmatter}
\title{Possible persistent current in a ring made of the perfect crystalline insulator}

\author[IEE]{A. Mo\v{s}kov\'{a}},
\author[IEE]{M. Mo\v{s}ko},
\ead{martin.mosko@savba.sk}
\author[IEE]{A. Gendiar}

\address[IEE]{Institute of Electrical Engineering, Slovak Academy of
Sciences, 84104~Bratislava, Slovakia}

\begin{abstract}
A mesoscopic conducting ring pierced by magnetic flux is known to
support the persistent electron current. Here we propose
possibility of the persistent current in the ring made of the
perfect crystalline insulator. We consider a ring-shaped lattice
of one-dimensional "atoms" with a single energy level. We express
the Bloch states in the lattice as a linear combination of atomic
orbitals. The discrete energy level splits into the energy band
which serves as a simple model of the valence band. We show that
the insulating ring (with the valence band fully filled by
electrons) supports a nonzero persistent current, because each
atomic orbital overlaps with its own tail when making one loop
around the ring. In the tight-binding limit only the neighboring
orbitals overlap. In that limit the persistent current at full
filling becomes zero which is a standard result.
\end{abstract}

\begin{keyword}
one-dimensional transport \sep mesoscopic ring \sep persistent
current \sep coherence

\PACS 73.23.-b \sep 73.61.Ey
\end{keyword}
\end{frontmatter}

%

\begin{figure}[!hbtp]
\centerline{\includegraphics[clip,width=0.82\columnwidth]{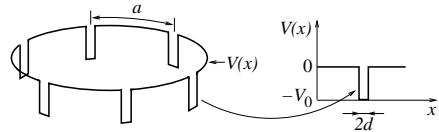}}
\vspace{-0.15cm} \caption{Potential $V(x)$ in a ring-shaped
periodic 1D lattice of atoms modelled by one-dimensional potential
wells of width $2d$ and depth $V_0$. The lattice period is $a$.
The ring circumference is $L = Na$, where $N$ is the number of
atoms.} \label{Fig:1}
\end{figure}

A mesoscopic conducting ring pierced by magnetic flux $\phi$ is
known to support the equilibrium persistent current.
\cite{Cheung,Levy}. Here we propose possibility of the persistent
current in a ring made of the perfect crystalline insulator like
for instance the intrinsic silicon crystal. Consider the
ring-shaped lattice of one-dimensional (1D) "atoms" in Fig.~\ref{Fig:1}.
The electron wave function $\psi(x)$ in such ring
obeys the Schr\"odinger equation
\begin{equation}
\label{schr rov bez mag} \left[-\frac
{\hbar^2}{2m}\frac{d^2}{dx^2} + V(x)\right]\psi(x) \ = \ E\psi(x)
\
\end{equation}
with the cyclic boundary condition \cite{Cheung}
\begin{equation}
\label{nova cyklicka} \psi(x+L) = \exp(i2\pi\phi/\phi_0)\psi(x) \
,
\end{equation}
where $m$ is the free electron mass, $x$ is the position along the
ring, $L$ is the ring length, $\phi_0 = h/e$ is the flux quantum,
and $V(x)$ is the potential of the lattice. The persistent current
is $I= -\partial \sum_n E_n(\phi)/\partial \phi$, where $\sum_n
E_n$ is the ground-state energy of all $N_e$ electrons in the
ring. Solving equation \eqref{schr rov bez mag} in the lattice
model with nearest neighbor hopping one gets \cite{Cheung} for odd
$N_e$
\begin{equation}
\label{perzistentny vysledok_tight_binding} I = -\frac {4\pi}{N
\phi_0} U \frac{\sin{\frac\pi N  N_e}}{\sin{\frac\pi N }}
\sin{\left(\frac{2\pi}{N}\frac{\phi}{\phi_0}\right)}
 \ ,
\end{equation}
where $U$ is the hopping amplitude. The persistent current
\eqref{perzistentny vysledok_tight_binding} is nonzero in the
conductor ($N_e<N$) but zero in the insulator ($N_e=N$). We want
to show that the persistent current is nonzero also in the
insulator.

The atomic orbital $\varphi(x)$ and energy $E^{at}$ in a single
isolated atom at $x=0$ obey the Schr\"odinger equation
\begin{equation}
\label{schr rov atomarna} \left[-\frac
{\hbar^2}{2m}\frac{d^2}{dx^2} + v(x)\right]\varphi(x) \ = \
E^{at}\varphi(x) \ ,
\end{equation}
where $v(x)$ is the atomic potential - the same potential well as
in Fig.~\ref{Fig:1} but with the infinitely thick barriers.

\begin{figure}[t]
\centerline{\includegraphics[clip,width=0.82\columnwidth]{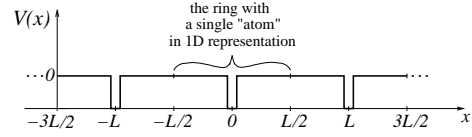}}
\vspace{-0.15cm} \caption{Potential $V(x)$ in the 1D ring with a
single 1D atom. The ring circumference $L$ is expanded as a linear
cell between $-L/2$ and $L/2$. The cell is then periodically
repeated, which is justified by the periodic condition \eqref{cykl
pot}.} \label{Fig:2}
\end{figure}

We want to express $\psi(x)$ via the isolated atomic orbitals. We
start with a toy model represented by the "single-atomic" ring in
Fig.~\ref{Fig:2}. In the ring geometry
\begin{equation}
\label{cykl pot} V(x+L)=V(x) \ .
\end{equation}
Due to the condition \eqref{cykl pot}, the ring potential $V(x)$
is formally tractable as an infinitely long periodic potential
with period $L$ (Fig.~\ref{Fig:2}). Therefore, $V(x)$ can be
expressed as an infinite sum of isolated atomic potentials,
\begin{equation}
\label{jednouzlovy potencial} V(x)\ = \ \sum_{j=-\infty}^{\infty}\
v(x-jL) \ ,
\end{equation}
albeit we treat the ring of the finite length $L$, not the
infinite crystal. Representation of the infinite crystal allows us
to expand $\psi(x)$ into the atomic orbitals as
\begin{equation}
\label{vlnova funkcia  jeden uzol} \psi(x) \ = \
\sum_{j=-\infty}^{\infty}\ e^{ikjL}\varphi(x-jL) \ .
\end{equation}
Expression \eqref{vlnova funkcia  jeden uzol} can be seen to obey
the boundary condition \eqref{nova cyklicka} for
$k=\frac{2\pi}{L}(\frac{\phi}{\phi_0}+n)$, where $n$ is the
integer.  We set $k=\frac{2\pi}{L}(\frac{\phi}{\phi_0}+n)$ into
the equation \eqref{vlnova funkcia  jeden uzol}. We get
\begin{equation}
\label{jedina vlnova funkcia  jeden uzol}  \psi(x) =
\sum_{j=-\infty}^{\infty}\ e^{i 2 \pi \frac{\Phi}{\Phi_0}
j}\varphi(x-jL) \ .
\end{equation}

We can now evaluate $\langle \psi\vert\psi\rangle$ as
\begin{eqnarray}
\label{pocitanie normy} \nonumber &&\langle \psi\vert\psi\rangle=
\int_{-\frac L2}^{\frac L2}dx \psi^{\ast}(x) \psi(x) =
\\
&& \sum_{j=-\infty}^{\infty} \sum_{j^{\prime}=-\infty}^{\infty}
e^{i 2\pi\frac {\phi}{\phi_0}(j-j^{\prime})} \int_{-\frac
L2}^{\frac L2}dx \varphi(x-j^{\prime}L)\varphi(x-jL) \nonumber
\\
&& = \sum_{\Delta_j=-\infty}^{\infty}e^{-i 2\pi\frac
{\phi}{\phi_0}\Delta_j}
\int_{-\infty}^{\infty}dx\varphi(x-\Delta_j L)\varphi(x) \ ,
\end{eqnarray}
where the integral in the second line was rewritten as
$\int_{-\frac L2-jL}^{\frac L2-jL}dx
\varphi(x-[j^{\prime}-j]L)\varphi(x)$, the variables $j$ and
$j^{\prime}$ were changed to the variables $j$ and $\Delta_j
\equiv j^{\prime}-j$, and the summation over $j$ was performed.

Further, we write the Hamiltonian in equation (\ref{schr rov bez
mag}) in the form $\hat H = \hat{H}^{at}_j + {V^{\prime}}(x-jL)$,
where
\begin{equation}
\label{atomarny hamiltonian_jL} \hat{H}^{at}_j \ = -\frac
{\hbar^2}{2m}\frac{d^2}{dx^2} + v(x-jL)
\end{equation}
%
is the Hamiltonian in \eqref{schr rov atomarna} for the atom at
site $jL$ and
\begin{equation} \label{atomarny potential_jL_reduced}
{V^{\prime}}(x-jL)=V(x)-v(x-jL) \ .
\end{equation}
For the electron energy $E =  \langle\psi\vert \hat H \vert
\psi\rangle / \langle \psi\vert\psi\rangle$ we obtain
\begin{eqnarray} \label{pomoc1}
\nonumber && E =  E^{at} + \frac 1{\langle\psi\vert\psi\rangle}
\sum_{j=-\infty}^{\infty} \sum_{j^{\prime}=-\infty}^{\infty} e^{i
2\pi\frac {\phi}{\phi_0}(j-j^{\prime})}
\\ && \times  \int_{-\frac L2}^{\frac L2}dx \varphi(x-j^{\prime}L)
{V^{\prime}}(x-jL) \varphi(x-jL) \ ,
\end{eqnarray}
where we have applied $\hat{H}^{at}_j\varphi(x-jL) = E^{at}
\varphi(x-jL)$. Using similar algebra as in equation
\eqref{pocitanie normy} we obtain
\begin{equation} \label{energy-1atom-exact}
E=E^{at} -\gamma_{0} - \frac 2{\langle\psi\vert\psi\rangle}
\sum_{\Delta_j=1}^{\infty} \gamma_{\Delta_j}  \cos{ (2\pi\frac
{\phi}{\phi_0}\Delta_j)} \ ,
\end{equation}
where $\gamma_{\Delta_j}= -
\int_{-\infty}^{\infty}\varphi(x-\Delta_jL) {V^{\prime}}(x)
 \varphi(x)$ is the overlap integral and we have applied $\gamma_{-\Delta_j}=\gamma_{\Delta_j}$.

\begin{figure}[t]
\centerline{\includegraphics[clip,width=\columnwidth]{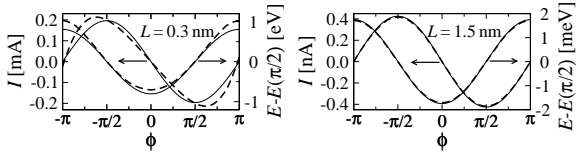}}
\vspace{-0.15cm} \caption{Ground-state energy and persistent
current versus magnetic flux for the "single-atomic" ring from
Fig.~\ref{Fig:2}. Two different ring lengths $L$ are considered.
The "atomic" parameters are $V_0 = 7.03$~eV and $2d=0.14$~nm,
they give a single atomic level $E_{at}=-3$~eV. The
formulae \eqref{vysledok pre zakladnu energiu} and \eqref{perz
prud pre jeden uzol} (shown in a full line) are compared with the
purely numerical approach (shown in a dashed line).} \label{Fig:3}
\end{figure}

We notice that $\gamma_{\Delta_j} \ll \gamma_{1}$  for
$|\Delta_j|>1$ and we also put $\langle \psi\vert\psi\rangle
\simeq 1$. The formula \eqref{energy-1atom-exact} thus reduces to
\begin{equation}
\label{vysledok pre zakladnu energiu} E  = E^{at} - \gamma_{0} -
2\gamma_1\cos{(2\pi\phi/\phi_0)} \  .
\end{equation}
We set \eqref{vysledok pre zakladnu energiu} into the expression
for the single-electron persistent current, $I= -\partial
E(\phi)/\partial \phi$. We get
\begin{equation}
\label{perz prud pre jeden uzol} I = -
(4\pi/\phi_0)\gamma_1\sin{(2\pi\phi/\phi_0)} \  .
\end{equation}
In Fig.~\ref{Fig:3} we compare the formulae \eqref{vysledok pre
zakladnu energiu} and \eqref{perz prud pre jeden uzol} with the
results obtained by solving the equation \eqref{schr rov bez mag}
numerically (the numerical method is described in \cite{Nemeth}).
Indeed, the formulae \eqref{vysledok pre zakladnu energiu} and
\eqref{perz prud pre jeden uzol} work well. The current
\eqref{perz prud pre jeden uzol} is nonzero due to the nonzero
overlap integral $\gamma_{1}$; the atomic orbital overlaps with
its own tail making one loop around the ring. In terms of hopping,
the electron hops around the ring back into its starting site.

In the multi-atomic ring in Fig.~\ref{Fig:1} the cyclic
condition \eqref{nova cyklicka} holds together with the Bloch
condition:
\begin{equation}
\label{periodicka Bloch vln funkcia N uzlov} \psi(x+L) =
e^{i2\pi\phi/\phi_0}\psi(x) \ \ \ , \ \ \ \psi(x+a) =
e^{ika}\psi(x).
\end{equation}
The ring potential in Fig.~\ref{Fig:1} is formally tractable as
an infinite lattice potential with period $a$, given as $V(x) =
\sum_{j=-\infty}^{\infty} v(x-ja)$. This allows to expand
$\psi(x)$ like
%
\begin{equation}
\label{vlnova funkcia  N uzol} \psi_k(x) \ = \
\sum_{j=-\infty}^{\infty}\ e^{ikja}\varphi(x-ja) \ .
\end{equation}
Expansion \eqref{vlnova funkcia  N uzol} obeys the boundary
conditions \eqref{periodicka Bloch vln funkcia N uzlov} for $k \ =
\frac{2\pi}{Na}\ \left( \frac{\phi}{\phi_0}+n\right)$, where $n$
is the integer.

Proceeding in analogy with equation \eqref{pocitanie normy} we
obtain
\begin{eqnarray}
\label{pocitanie normy N uzlov} \nonumber &&\langle
\psi_{k_n}\vert\psi_{k_n}\rangle= \int_{-\frac L2}^{\frac L2}dx
\psi_{k_n}^{\ast}(x) \psi_{k_n}(x) =
\\
&& \sum_{j=-\infty}^{\infty} \sum_{j^{\prime}=-\infty}^{\infty}
e^{i k_n(j-j^{\prime})a} \int_{-\frac L2}^{\frac L2}dx
\varphi(x-j^{\prime}a)\varphi(x-ja) \nonumber
\\
&& = N \sum_{\Delta_j=-\infty}^{\infty}e^{-i k_n \Delta_j a}
\int_{-\infty}^{\infty}dx\varphi(x-\Delta_j a)\varphi(x) \ ,
\end{eqnarray}
where the integral in the second line was rewritten as
$\int_{-\frac L2-ja}^{\frac L2-ja}dx
\varphi(x-[j^{\prime}-j]a)\varphi(x)$ and we summed over $j$.

We write the Hamiltonian in equation (\ref{schr rov bez mag}) as
$\hat H = \hat{H}^{at}_j + {V^{\prime}}(x-ja)$, where
$\hat{H}^{at}_j \ = -\frac {\hbar^2}{2m}\frac{d^2}{dx^2} +
v(x-ja)$ and $V^{\prime}(x-ja)=V(x)-v(x-ja)$. Using the above
relations we express $E_n = \langle\psi_{k_n}\vert \hat H \vert
\psi_{k_n}\rangle / \langle \psi_{k_n}\vert\psi_{k_n}\rangle$ as
\begin{eqnarray} \label{pomoc1N}
\nonumber && E_n =  E^{at} + \frac
1{\langle\psi_{k_n}\vert\psi_{k_n}\rangle}
\sum_{j=-\infty}^{\infty} \sum_{j^{\prime}=-\infty}^{\infty} e^{i
k_n(j-j^{\prime})a}
\\ && \times  \int_{-\frac L2}^{\frac L2}dx \varphi(x-j^{\prime}a)
{V^{\prime}}(x-ja) \varphi(x-ja) \ ,
\end{eqnarray}
where we have applied $\hat{H}^{at}_j\varphi(x-ja) = E^{at}
\varphi(x-ja)$. Using similar manipulation as in equation
\eqref{pocitanie normy N uzlov} and approximating
$\langle\psi_{k_n}\vert\psi_{k_n}\rangle \simeq N$, we obtain
\begin{equation} \label{energy-Natom-exact}
E_n=E^{at} - \gamma_{0} - 2 \sum_{\Delta_j=1}^{\infty}
\gamma_{\Delta_j} \cos{ (k_n \Delta_j a) } \ ,
\end{equation}
where $\gamma_{\Delta_j}= -
\int_{-\infty}^{\infty}\varphi(x-\Delta_ja) {V^{\prime}}(x)
 \varphi(x)$ is the overlap integral and  $\gamma_{-\Delta_j}=\gamma_{\Delta_j}$.
Setting \eqref{energy-Natom-exact} into the single-electron
current $I_n= -\partial E_n/\partial \phi$ we obtain
\begin{equation}
\label{perz prud N uzlov pre stav n} I_n \ = \ - \ \frac
{4\pi}{\phi_0} \sum_{\Delta_j=1}^{\infty} \frac{\Delta_j}{N}
\gamma_{\Delta_j}\sin{(k_n\Delta_j a)} \  .
\end{equation}
We sum $I = \sum_n I_n$ over $n=0, \pm 1, \pm 2, \dots \pm
(N_e-1)/2$ assuming  odd $N_e$ and spinless system. For $N_e = N$
\begin{equation}
\label{perzistentny full} I  = -(4\pi/\phi_0) N \gamma_N \
\sin(2\pi\phi/\phi_0) \ ,
\end{equation}
which the persistent current in the insulating 1D ring.
We note that if we keep in~\eqref{perz prud N uzlov pre stav n}
only the term with $\Delta_j=1$, the sum $I=\sum_n I_n$ gives
the standard formula \eqref{perzistentny vysledok_tight_binding}.

\begin{figure}[t]
\centerline{\includegraphics[clip,width=\columnwidth]{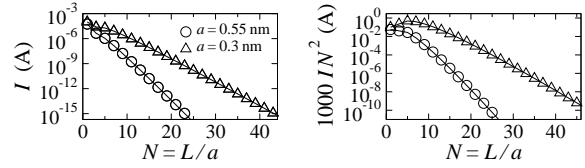}}
\vspace{-0.15cm} \caption{Left panel: Persistent current $I$ versus
ring length $L$ for the 1D ring made of the perfect crystalline
insulator, calculated using the formula \eqref{perzistentny full}.
THe right panel shows the same results multiplied by $N^2$ to estimate
the persistent current in a hollow 3D cylinder with cross section
of order $\sim L^2$ and circumference $\sim L$. The factor of 1000
appears when $10^6$ cylinders are measured together. Here we have
used the parameters $V_0=1.74$~eV, $2d=0.01$~nm (giving a single
atomic level $E_{at}=-0.2$~eV) and two different $a$.} \label{Fig:4}
\end{figure}

Fig.~\ref{Fig:4} shows the persistent current \eqref{perzistentny full}
as a function of $L$. It decays with $L$ exponentially. Enhancement
by factor $\sim N^2$ can be achieved using the 3D cylinder and by
factor of $10^3$ using $10^6$ cylinders on a single chip~\cite{Levy}.
But even the largest considered $L$ is not achievable to present
nanotechnology. Due to the 1D atoms, our model is a minimum model.
It shows that the persistent current in the insulator exists but
it strongly underestimates the effect. For realistic 3D insulators
we expect larger persistent currents and technologically feasible
sample dimensions. Another possibility is to fabricate the ring-shaped
1D Kronig-Penney model at full filling which also shows the
persistent current~\cite{NemethMosko}.

We thank for the APVV grant APVV-51-003505.

%
%

\end{document}